\documentclass[12pt,]{article}
\usepackage{authblk}
\usepackage{fullpage}
\usepackage[utf8x]{inputenc}
\usepackage[T1]{fontenc}
\usepackage{siunitx}
\usepackage[version=3]{mhchem}
\usepackage{mathptmx}
\usepackage{pifont}

\usepackage{enumerate}
\usepackage{amsthm}
\usepackage{amsmath}                  
\usepackage{amssymb}                 
\usepackage{amsfonts}
\usepackage{array,color}
\usepackage{xcolor}
\usepackage{extarrows}
\usepackage{cases}
\usepackage{mathrsfs}                 
\usepackage[sectionbib]{natbib}
\usepackage{graphicx}
\usepackage{float}
\usepackage{subfloat}
\usepackage{subfigure}
\usepackage{bm}  
\usepackage{diagbox}
\usepackage{threeparttable}
\usepackage{rotating}

\usepackage[left]{lineno}
\nolinenumbers

\usepackage{setspace}
\usepackage[unicode=true]{hyperref}
\hypersetup{breaklinks=true,
            bookmarks=true,
            colorlinks=false,
            pdfborder={0 0 0}}
\usepackage{graphicx}
\usepackage{multirow}
\graphicspath{{figures/}}

 \usepackage{bm}   
 \newtheorem{theorem}{Theorem}

 \newtheorem{assumption}{Assumption}

\makeatletter
\def\ScaleWidthIfNeeded{%
 \ifdim\Gin@nat@width>\linewidth
    \linewidth
  \else
    \Gin@nat@width
  \fi
}
\def\ScaleHeightIfNeeded{%
  \ifdim\Gin@nat@height>0.9\textheight
    0.9\textheight
  \else
    \Gin@nat@width
  \fi
}
\makeatother
\setkeys{Gin}{width=\ScaleWidthIfNeeded,height=\ScaleHeightIfNeeded,keepaspectratio}%

\title{A Stable and Adaptive Polygenic Signal Detection Method Based on Repeated Sample Splitting}
\author[1]{Yanyan Zhao}
\author[1,2]{Lei Sun}
\affil[1]{Department of Statistical Sciences, University of Toronto, 100 St. George Street, Toronto, Ontario M5S 3G3, Canada\\ 
	
yanyan.zhao@utoronto.ca}

\affil[2]{Division of Biostatistics, Dalla Lana School of Public Health, University of Toronto,
155 College Street, Toronto, Ontario M5T 3M7, Canada\\ lei.sun@utoronto.ca}

\begin{document}

\maketitle

\section*{Abstract}
	Focusing on polygenic signal detection in high dimensional genetic association studies of complex traits, we develop an adaptive test for generalized linear models to accommodate different alternatives. To facilitate valid post-selection inference for high dimensional data, our study here adheres to the original sampling-splitting principle but does so, repeatedly, to increase stability of the inference. We show the asymptotic null distributions of  the proposed test for both fixed and diverging number of variants. We also show the asymptotic properties of the proposed test under local alternatives, providing insights on why power gain attributed to variable selection and weighting can compensate for efficiency loss due to sample splitting.  We support our analytical findings through extensive simulation studies and two applications. The proposed procedure is computationally efficient and has been implemented as the R package DoubleCauchy.
	   \\[0.5cm]
{\it Keywords}: Adaptive; High dimensional data; Polygenic risk score; Robustness; Sample splitting.

\section{Introduction}\label{intro}

Polygenic signal detection can improve power of genetic studies of complex traits by aggregating weak signals across a large number of genetic variants that do not, individually, achieve statistical significance. The general concept of set-based testing has been well examined in settings such as gene-based association studies  \citep{Derkach2014, Zhao2020} or multiple-phenotype analyses \citep{Liu2018}, but applications of existing methods to high dimensional genetic data require additional considerations. 

For simultaneously testing regression coefficients in high dimensional generalized linear models (GLMs), \citet{Goeman2011} proposed a feasible test statistic for the scenario when the number of variants is fixed but can be larger than the sample size. \citet{Guo2016} first investigated the asymptotic properties of the test statistic of \citet{Goeman2011} for diverging number of variants, and then proposed a U-statistic for GLMs with unbounded link functions. The p-value calculation based on asymptotical normal approximation, however, is not accurate for stringent significance levels, and the test is not adaptive to different alternatives. 

Recently, \citet{Wu2019} proposed an adaptive method, where the test statistic is based on different functions of variant-specific score statistics, with different functions targeting different alternatives. However, accurate inference requires parametric bootstrap, which is computationally expensive for large-scale studies or stringent significance levels. Further, the method of \citet{Wu2019} aggregates information across {\it all} variants, and the lack of variable selection can adversely affect power despite of its using the whole sample to derive the variant-specific score statistics. 

In this paper, we focus on polygenic signal detection in high dimensional generalized linear models, and we propose to use sample splitting, repeatedly, for valid and stable post-variable-selection inference. One sample splitting produces two independent sub-samples, of which one is used for variable selection, and the other for valid association testing without the need for correcting for variable-selection bias. This general principle has been used in many study settings, but the inherent instability has been noted, including in variable selection \citep{Larry2009, Meinshausen2009, Meinshausen2010}, change-point detection \citep{Zou2020}, and more recently selective inference \citep{Rinaldo2019, Barber2019, Dai2020}. Repeated sample splitting is a natural remedy, but it is not obvious how to aggregate information across multiple, correlated sample splits to derive a valid and efficient test. 

In the context of polygenic signal detection, the first polygenic risk score (PRS) method \citep{ISC2009} used the one-time-only sample splitting strategy. Specifically, the method divides the data into a training sample and a testing sample, and then performs a two-stage analysis. Stage one applies a variable selection procedure to the training sample to obtain a set of potentially associated variants and their corresponding weights. Using the independent testing sample, stage two first constructs a polygenic risk score for each individual by calculating a weighted sum of the numbers of the risk allele across the selected variants, and it then evaluates the aggregated score for association with the trait of interest. The original polygenic method has since been extended \citep{Vilhj2015,Shi2016,Lloyd2019,Li2020}, but the strategy of repeated sample splitting has not been examined.   

In this work, we combine the concepts of ``repeated sample splitting" and ``adaptive testing" together to develop a robust polygenic association test for testing high dimensional regression coefficients in generalized linear models. In Section \ref{method}, we first review the classical polygenic association test, based on a weighted sum of the numbers of the risk allele across the selected variants, and we note its equivalence to a weighted sum of the variant-specific score statistics. We then consider different weighting factors for the score vector, where the different weights are tailored to different alternatives. To aggregate information across the different weighting factors, we use the recent Cauchy method of \citet{Liu2019}, and we discuss the connection of our adaptive method with that of \citet{Wu2019}. To improve stability of our inference, we then introduce the combination procedure that aggregates information across multiple, correlated sample splits. Finally, we derive the asymptotic null distributions of the proposed test for both fixed and diverging number of variants, and we study its asymptotic properties under local alternatives. In Section \ref{simulation} we present extensive simulation results for method evaluation and comparison. In Section \ref{application} we provide results from two applications, including additional simulation studies using the real genetic data from the two applications combined with simulated outcome data. We conclude with discussion in Section \ref{discussion}, which includes information for DoubleCauchy, a R package that implements the proposed test. 

\section{Methods}\label{method}  

\subsection{Notations}\label{method_notations}

Let $Y \in \mathbb{R}^{n \times 1}$ be the outcome variable of interest, $G \in \mathbb{R}^{n \times J}$ the genotype matrix, and $ X \in \mathbb{R}^{n \times q} $ the covariate matrix for a sample of size $n$ with $J$ genetic variants and $q$ covariates. For clarity, let $y_i$ be the response for individual $i$,  $g_{ij}$ the genotype for individual $i$ and variant $j$, and $x_{ij'}$ the covariate value for individual $i$ and covariate $j'$, $i=1,\ldots, n$, $j=1,\ldots, J$, and $j'=1,\ldots, q$. Further, 
let $G_{i} \in \mathbb{R}^{J \times 1}$ be the genotype vector for individual $i$, $G_{j} \in \mathbb{R}^{n \times 1} $ the genotype vector for variant $j$, $ X_{i} \in \mathbb{R}^{q \times 1}$ the covariate vector for individual $i$, and $ X_{j'} \in \mathbb{R}^{n \times 1}$ the vector for covariate $j'$. 

We assume that conditional on $(G_i, X_i)$, $y_i$ follows a distribution with density function, $f(y_i)=exp\{(y_i\theta_i-b(\theta_i))/a_i(\phi)+c(y_i,\phi)\}$ for some specific functions $a(\cdot)$, $b(\cdot)$ and $c(\cdot)$, where $\theta$ is the canonical parameter, $\phi$ the dispersion parameter, $\text{var}(y_i| G_i, X_i) = a_i(\phi)\nu(\mu_i)$, and $\nu(\mu_i)$ the variance function.  We consider the generalized linear model that models $\mu_i=b'(\theta_i)=E(y_i| G_i, X_i)$ for different types of response variables in the exponential family by a monotone and differentiable link function $\mathcal{G}(\cdot) $, $\mathcal{G}(\mu_i) =  G_i^T \beta + X_i^T\beta_x$, where $ \beta$ and $ \beta_x$ are, respectively, $J$- and $q$-dimensional vectors of regression coefficients; $q$ is fixed but $J$ may vary depending on the study setting. Among the $J$ genetic variables, we use $\mathcal{M}^*$ and $|\mathcal{M}^*|$ to denote, respectively, the set and number of truly associated ones. For simplicity but without loss of generality, we also assume that $G_i$ and $X_i$ have been mean centred at zero and standardized to have variance one. 

\subsection{The classical polygenic risk score for high dimensional association test}\label{method_prs}

Suppose we have $2n$ independent observations, the classical polygenic risk score-based high dimensional association testing method \citep{ISC2009} first randomly splits the sample into two equal subsets, $D_{n,1}$ and $D_{n,2}$; the corresponding data and parameter estimates such as $y$, $X$, $G$,  and $\hat \beta$ will carry superscripts $^{(1)}$ and $^{(2)}$, respectively for the two subsets, unless specified otherwise.  A variable selection procedure is then applied to the training sample $D_{n,1}$ to select a subset of candidate genetic variables, $\mathcal{M}$, where we define $J_2 = |\mathcal{M}|$. 

To test the $J_2$ variants simultaneously in the testing sample $D_{n,2}$, one single polygenic risk score $G_i^*$ is constructed by aggregating the $J_2$ selected variables using $G^{(2)}_{i}$, but weighted by the effect estimate $\hat{\beta}^{(1)}_j$ obtained from $D_{n,1}$,  $j \in \mathcal{M}$ . That is, $G_i^* =  \sum_{j=1}^{J_2} \hat{\beta}^{(1)}_jg^{(2)}_{ij}, i=1, \ldots, n$.  
The inference is then based on the generalized linear regression model applied to $D_{n,2}$, 
\begin{equation}\label{eq:1}
	\mathcal{G}\{E(y^{(2)}_i|G^*_i, X_i^{(2)})\} =  G_i^*\beta^* + {X_i^{(2)T}}\beta_x, 
\end{equation}
and testing 
\begin{equation}\label{h0s}
	H_0: \beta^*=0 \:\: \text{versus} \:\: H_1: \beta^* \neq 0.    
\end{equation}
The corresponding score statistic is,  
$T_1  = \sum_{i=1}^n (y^{(2)}_i-\hat{\mu}_{i}^{(2)})G_i^*$,
where $\hat{\mu}_{i}^{(2)} = \mathcal{G}^{-1}({X_i^{(2)T}}\hat \beta_{x})$ and $\hat{\beta}_{x}$ is the maximum likelihood estimate of $\beta_x$ under $H_0$. The distribution of standardized $T_1$ can be approximated by $\chi_1^2$, and the p-value of a test based on $T_1$ will be denoted as $p_1$. 

This classical polygenic association testing has since been improved on several fronts, including modelling dependency structure (i.e.\ linkage disequilibrium) between genetic variables \citep{Vilhj2015} and better estimation of $\beta_j^{(1)}$ \citep{Shi2016}, among others \citep{Lloyd2019}. However, additional work are needed. To facilitate our discussion, first it is instructive to re-formulate $T_1$ as the following,
\begin{align*} 
	T_1 & = \sum_{i=1}^n (y^{(2)}_i-\hat{\mu}_{i}^{(2)})G_i^* = \sum_{i=1}^n (y^{(2)}_i-\hat{\mu}_{i}^{(2)})\sum_{j=1}^{J_2} \hat{\beta}^{(1)}_j g^{(2)}_{ij} \\
	&= \sum_{j=1}^{J_2} \hat{\beta}^{(1)}_j \sum_{i=1}^n (y^{(2)}_i-\hat{\mu}_{i}^{(2)}) g^{(2)}_{ij} = n\sum_{j=1}^{J_2} \hat{\beta}^{(1)}_j S_j,
\end{align*}            
where $S_j = n^{-1}\sum_{i=1}^n (y^{(2)}_i-\hat{\mu}_{i}^{(2)}) g^{(2)}_{ij}$. 
Thus, $T_1$ constructed based on the aggregated risk score $G_i^*$ is analytically equivalent to a {\it linearly} weighted average of the score statistics, $S_j$'s, across the $J_2$ genetic variants. 

\begin{figure}[!htp]
	\centering
	\includegraphics[width=0.75\textwidth]{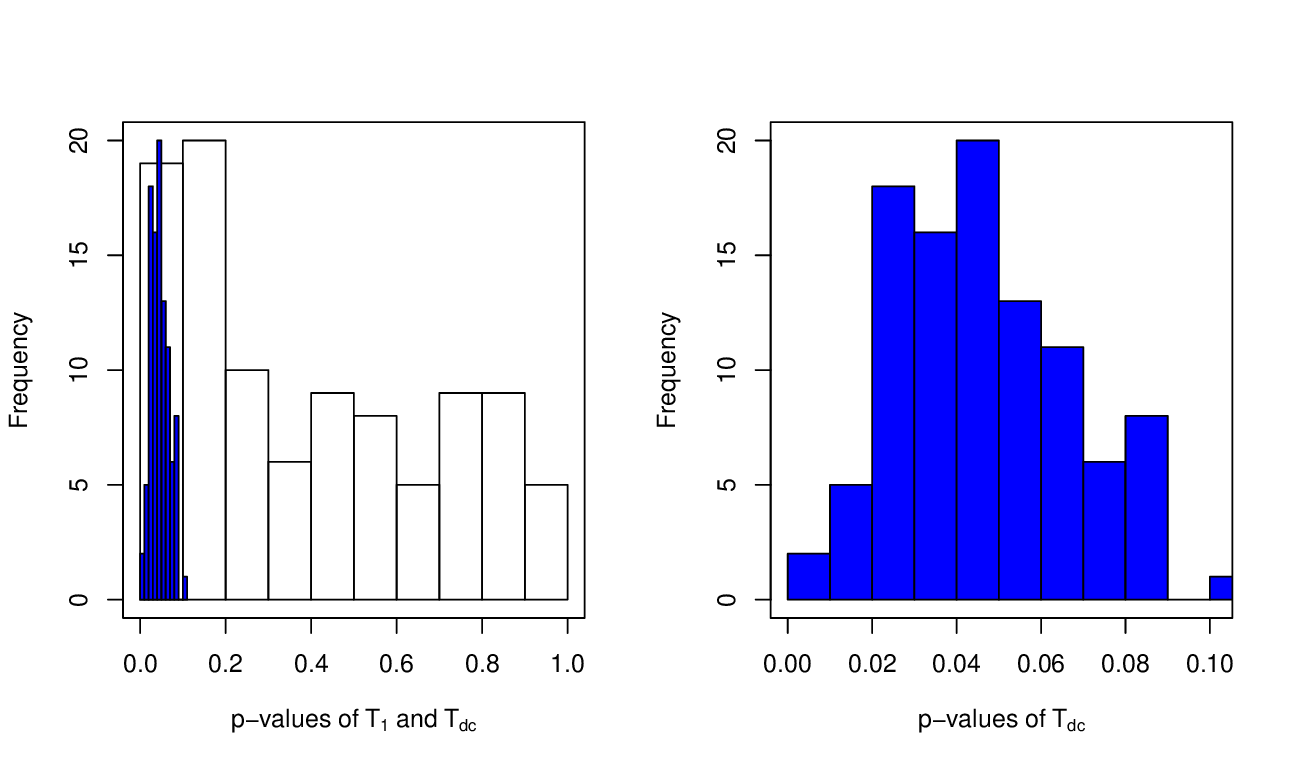}
	\caption{\label{fig:cf} Histograms of p-values of $T_1$ (white) and the proposed $T_{dc}$ (blue) based on randomly splitting a real dataset to training and testing samples, independently 100 times.  The right figure is a zoom-in plot for the proposed $T_{dc}$. Details of the application data in Section \ref{application}.}
\end{figure}

Tests based on $T_1$ are sub-optimal when signs of $\hat{\beta}^{(1)}_j$ and $S_j$ differ. When the effect size $\beta_j$ is large, it is likely to obtain sign-consistent results between $\hat{\beta}^{(1)}_j$ from the training sample and $S_j$ from the testing sample. This will prevent $S_j$'s of variants with opposite direction of effect being cancelled out. However, for weak signals there is no theoretical guarantee for obtaining sign-consistent $\hat{\beta}^{(1)}_j$ and $S_j$ \citep{Jin2014}, so it is better to develop a test that is robust to this assumption. Recent work in association tests for rare variants have also shown that $T_1$ type of tests are only powerful when a large proportion of the variants being tested are causal, in addition to their genetic effects being in the same direction \citep{Derkach2014}.  Further, the direct use of $\hat{\beta}^{(1)}_j$'s as weights may not be robust to different alternatives. Finally, when the signal-to-noise ratio is low as often the case in practice, the one-time-only sample splitting approach may not be reliable \citep{Meinshausen2009}. Figure \ref{fig:cf} is an illustration of the p-value lottery phenomenon associated with $T_1$, when it is applied to a real dataset with $2n=1409$ and $J=3754$; see Section \ref{application} for details of the application data. 

\subsection{An adaptive procedure for polygenic signal detection}\label{method_adaptive}

Here we develop a robust method that is adaptive to different alternatives. We first propose new tests by considering different weighting schemes, given a particular sample split. We then improve the stability of our inference through repeated sample splitting. 

Recall that testing (\ref{h0s}) in (\ref{eq:1}) can be reformulated as testing 
\begin{equation}\label{h0}
	H_0: \beta=0 \:\:  \text{versus} \:\: H_1: \beta \neq 0, 
\end{equation} 
in 
\begin{equation}\label{eq:2}
	\mathcal{G}\{E(y^{(2)}_i|G_i^{(2)}, X_i^{(2)})\} =  G_i^{(2)T}\beta + {X_i^{(2)T}}\beta_x,
\end{equation}
where $\beta=(\beta_1,\ldots, \beta_{J_2})^T$. The proposed new test statistics have the following form,
\begin{equation}\label{eq:Tgamma}
	T_{\gamma} = n \sum_{j=1}^{J_2} {w}_j^{\gamma-2} S_j^2,\:\:  \gamma \in \Gamma = \{2, 4, 6, \ldots\}, 
\end{equation}
where $w_j$ depends on $\hat{\beta}^{(1)}_j$ obtained from $D_{n,1}$, and $\gamma$ is an even integer to avoid signal cancellation between variants.

Let $S=(S_1, \ldots, S_{J_2})^T$ and $R=\text{diag}\{r_j\}=\text{diag}\{{w}_j^{\gamma-2}\}$, $j=1,\ldots,J_2$, we have 
\begin{equation}\label{eq:3}
	T_{\gamma} = n{S^{T}RS}.   
\end{equation}
We can easily modify $R$ to include off-diagonal elements to reflect potential linkage disequilibrium between genetic variables, and we will study the asymptotic null distributions of $T_{\gamma}$  in Theorems 1--3 for both fixed and diverging $J_2$.

The different $\gamma$ values in (\ref{eq:Tgamma}) adapt to different signal sparsities. To obtain an accurate yet computationally efficient adaptive test, we propose to aggregate $p_{\gamma}$'s, the p-values of $T_{\gamma}, \gamma \in \Gamma= \{2, 4, 6, \ldots\}$, and $p_1$, the p-value of $T_1$, using the Cauchy combination method recently proposed by \citet{Liu2019}. The Cauchy method can accommodate complex dependency structure among p-values without explicitly modelling it. In our setting, the proposed test statistics is 
\begin{equation}\label{eq:Tc}
	T_{c}={(|\Gamma|+1)}^{-1} \sum_{\gamma \in \Gamma \cup {1}} \tan\{(0.5-p_{\gamma})\pi\}.
\end{equation}
The tail of the null distribution of $T_{c}$ can be well approximated by the standard Cauchy distribution, as long as the individual $p_{\gamma}$'s are accurate, which we study in Sections \ref{method_asymptotic} and \ref{simulation}. The final p-value of $T_{c}$ is $p_c=1/2-(\text{arctan}\ t_{c})/\pi$, where $t_{c}$ is the observed value for $T_c$. 

Here we acknowledge that $T_{\gamma}$ is related to SPU type of test statistics proposed by \citet{Wu2019}. For an integer $\gamma \geq 1$, 
$\text{SPU}(\gamma) = \sum_{j=1}^J S_j^{\gamma},$
where $S_j$ is obtained from the {\it whole} sample. If we omit the sample splitting step in our approach, $J_2=J$, and let $w_j = S_j$, we have $T_{\gamma} \propto \text{SPU}(\gamma)$ for all $\gamma>1$. 
The authors of $\text{SPU}(\gamma)$ have noted that for an {\it even} integer $\gamma \rightarrow \infty$, $\text{SPU}(\gamma) \propto (\sum_j |S_j|^{\gamma})^{1/\gamma} \rightarrow \max_j|S_j|$, defined as $\text{SPU}(\infty)$; this suggests that larger $\gamma$ is more powerful for sparse alternatives. To make SPU robust to different alternatives, the authors then proposed an adaptive SPU, $\text{aSPU} = \text{min}_{\gamma \in \Gamma_{\text{aSPU}}} \{p_{\text{spu}(\gamma)}\},$ where the recommended $\Gamma_{\text{aSPU}} = \{1,2,3,4,5,6,\infty\}$, and $p_{\text{spu}(\gamma)}$ is the p-value of $\text{SPU}(\gamma)$. The asymptotic $p_{\text{spu}(\gamma)}$ for $\gamma=1$ and $2$ can be obtained with mild conditions imposed on moments of $S_j$'s and their correlation structure \citep{Wu2019}, but the asymptotic approximation is not accurate for $\gamma > 2$. The authors then proposed to calculate $p_{\text{spu}(\gamma)}$, and subsequently $p_{\text{aSPU}}$, based on parametric bootstrap, which is computational expensive. 

The distinction between $\text{aSPU}$ and the proposed $T_c$ is four-fold. Firstly, although $T_c$ includes evidence from $T_1$, the building block of $T_c$ is $T_{\gamma}$, where $\gamma$ is an even integer, which facilitates studies of asymptotic properties of the proposed tests; see Theorems 1--4 for details. Secondly, tests using different $\gamma$  values are correlated with each other. Thus, even if the individual p-value estimation is accurate, the minimum-p approach of $\text{aSPU}$ makes the inference more difficult than that of $T_c$, which is based on the easy-to-implement Cauchy method. Thirdly, although $\text{aSPU}$ uses the whole sample for association testing, it aggregates information across all $J$ genetic variants, many of which may be from the null leading to reduced power as compared to $T_c$, which benefits from variable selection. Lastly, the flexible structure of $w_j$ in $T_{\gamma}$ can incorporate other information available for each variant $j$, such as the functional importance measure of a genetic variant \citep{Iuliana2016}.

To further robustify $T_{c}$ against sampling variation inherent in the one-time-only sample splitting approach, we then consider repeated sample splitting of $m$ times. For the $s$th sample split, $s =1, \ldots, m$, we obtain $T_{c,s}$ and its corresponding p-value, $p_{c,s}$. To combine the $p_{c,s}$'s while not explicitly modelling the correlation, we again utilize the Cauchy method of \citet{Liu2019}. The proposed double Cauchy combination test statistic is
\begin{equation}\label{eq:Tdc}
	T_{dc}=m^{-1} \sum_{s=1}^m \tan\{(0.5-p_{c,s})\pi\}.
\end{equation}
Similar to inference based on $T_c$, the tail of the null distribution of $T_{dc}$ can be well approximated by the standard Cauchy distribution, as long as the individual p-values to be combined are accurate which we study next.

\subsection{Asymptotic properties of $T_{\gamma}$}
\label{method_asymptotic}

To make the dependency of $T_{\gamma}$ on $n$ and $J_2$ explicit, we use $T_{n,J_2,\gamma}$ to denote $T_{\gamma}$ in this section. We study the asymptotic properties of $T_{n,J_2,\gamma}$ for both fixed and diverging $J_2$, under the null or local alternatives.  For notation simplicity, we now omit superscript $^{(2)}$ from $Y \in \mathbb{R}^{n \times 1}$, $G \in \mathbb{R}^{n \times J_2}$ and $X \in \mathbb{R}^{n \times q}$, representing, respectively, the outcome, genotype and covariate data in the testing sample $D_{n,2}$, where $J_2$ is the number of variants to be tested.  Recall that $T_{n,J_2,\gamma} = n{S^{T}RS}$, where $S=(S_1, \ldots, S_{J_2})^T$ is the score vector, $R=\text{diag}\{r_j\}$, and $\gamma$ is an even integer. The covariance matrix of $n^{1/2}S$ is  $\Sigma_s=E\{a_i(\phi)\nu(\mu_i)G_iG_i^T\}$, where $G_i \in \mathbb{R}^{ J_2 \times 1}$, the genotype vector for individual $i$, $\epsilon=(\epsilon_1, \ldots, \epsilon_n)^T=Y-\mathcal{G}^{-1}(G\beta+X\beta_x)$, and $\epsilon_0 = (\epsilon_{01}, \ldots, \epsilon_{0n})^T = Y-\mathcal{G}^{-1}(X\beta_x)$. 

The following theorem gives the asymptotic null distribution of $T_{n,J_2,\gamma}$, provided that the same regularity conditions, required for the convergence of $S$ to a multivariate normal random variable, hold \citep{Goeman2011}. In addition, we ignore the nuisance parameters $a_i(\phi)$ and $\beta_x$ for now and discuss how to include them in Section \ref{discussion}. We provide all proofs in the Supplementary Material.

\begin{theorem}
	\label{thm1}
	Under the null hypothesis $H_0$ in (\ref{h0}), for any fixed finite $J_2$ and $\gamma$, $T_{n,J_2,\gamma} \rightarrow T_{J_2, \gamma}$ 
	in distribution as $n \rightarrow \infty$, where $T_{J_2,\gamma}$ and $\sum_{j=1}^{J_2} \lambda_{J_2,j}\chi_{1j}^2$ are equivalent in distribution, $\chi_{1j}^2$'s are independent variables with the central chi-square distribution with 1 degrees of freedom, $\chi_1^2$, $\lambda_{J_2,1} \geq \ldots, \geq \lambda_{J_2,J_2}$ are the eigenvalues of $C_s^TRC_s$, and $\Sigma_s = C_sC_s^T$.
\end{theorem}

When $Y$ is normally distributed, $T_{n,J_2,\gamma}$ and $T_{J_2,\gamma}$ equivalent in distribution always holds for any $n$ (and finite $J_2$); when both $n$ and $J_2$ are diverging, additional assumptions are required. 

\begin{assumption}
	\label{asm:1} 
	Assume $G_{i} = C_{g}Z_{i}, \forall i$, where $C_g$ is a $J_2 \times J_2$ matrix and $C_gC_g^T = \Sigma_g$, and $Z_i=(z_{i1}, \ldots, z_{iJ_2})^T$  with $E(Z_i) = 0$ and $\text{cov}(Z_i) = I_{J_2}$. Assume $z_{ij}$ has finite eighth moment and $E(z_{ij}^4) = 3 + \Delta \ < \infty$, $\forall j$, where $\Delta$ is a constant and $\Delta > -3$, and $E(\Pi_j z_{ij}^{\nu_j}) = \Pi_j E(z_{ij}^{\nu_j})$, where $ \sum_j \nu_j  \leq 8$ and all $\nu_j$'s are non-negative integers.
\end{assumption}
\begin{assumption}
	\label{asm2}
	Let $f_g$ be the probability density of $G$ and $D(f_g)$ be its support. Assume $E(\epsilon \mid G) = 0$ and $E(\epsilon^3 \mid G) = 0$, and there are positive constants $K_1$ and $K_2$ such that $E(\epsilon^2 \mid G) > K_1$ and $E(\epsilon^4 \mid G) < K_2$ almost everywhere for $g \in D(f_g)$.
\end{assumption}
\begin{assumption}
	\label{asm3}
	There exist real numbers $\rho_{\infty,j}$'s such that $\lim_{J_2 \rightarrow \infty} \rho_{J_2,j} = \rho_{\infty, j}$ uniformly $\forall j$, and $\lim_{J_2 \rightarrow \infty} \sum_{j=1}^{J_2} \rho_{J_2,j} = \sum_{j=1}^{\infty} \rho_{\infty,j} < \infty$, where $\rho_{J_2,j} = \lambda_{J_2,j}/\sqrt{tr(R\Sigma_s)^2},\ j=1, \ldots, J_2$, which are the eigenvalues of $C_s^TRC_s/\sqrt{tr(R\Sigma_s)^2}$ in descending order.
\end{assumption}
\begin{assumption}
	\label{asm4}
	$n \{tr(R\Sigma_g)^2/tr^2(R\Sigma_g)\} \rightarrow \infty$ as $n$ and $J_2 \rightarrow \infty$. 
\end{assumption}

Assumptions 1--3 are standard in studying high dimensional testing \citep{Guo2016,Zhang2019}. Assumption 4 specifies a relationship between $n$ and $J_2$. Because $tr(R\Sigma_g)= \sum_{j=1}^{J_2} \lambda_{J_2,j}$ and $tr(R\Sigma_g)^2 = \sum_{j=1}^{J_2} \lambda_{J_2,j}^2$, we have $n \{tr(R\Sigma_g)^2/tr^2(R\Sigma_g)\} = n\{\sum_{j=1}^{J_2} \lambda_{J_2,j}^2/(\sum_{j=1}^{J_2} \lambda_{J_2,j})^2\}$. Thus, Assumption 4 holds for any diverging $n$ and $J_2$ if $\lambda_{J_2,j}$'s are dominated by first few larger ones. When all $\lambda_{J_2,j}$'s are similar in magnitude, Assumption 4 is equivalent to requiring sample size $n$ grows to infinity at a rate faster than $J_2$. The following theorem generalizes Theorem 1 from finite to infinite $J_2$. 

\begin{theorem}
	\label{thm2}
	Under the null hypothesis $H_0$ in (\ref{h0}) and assume Assumptions 1--4 hold, 
	\[
	\sigma_{n,0}^{-1}\{ T_{n, J_2,\gamma} - tr(R \Sigma_s)\} \rightarrow \zeta\: \text{ and } \:  \{2tr(R \Sigma_s)^2\}^{-1/2}\{ T_{J_2,\gamma} - tr(R \Sigma_s)\} \rightarrow \zeta 
	\]
	in distribution as $n$ and $J_2 \rightarrow \infty$, where $\zeta$ and $\sum_{j=1}^{\infty} \rho_{\infty,j}(\chi_{1j}^2-1)/\sqrt{2}$ are equivalent in distribution, $\sigma_{n,0}^2= 2tr(R \Sigma_s)^2 + \delta$, and
	$\delta = n^{-1}\left\{\sum_{j=1}^{J_2}\sum_{k=1}^{J_2} r_jr_k E(g_{ij}^2g_{ik}^2\epsilon_{0i}^4)-tr^2(R\Sigma_s) - 2tr(R\Sigma_s)^2\right\}=o\{tr(R\Sigma_s)^2\}$. 
	Therefore, as $n$ and $J_2 \rightarrow \infty$, 
	\[
	\text{sup}_x|pr(T_{n,J_2,\gamma} \leq x) - pr(T_{J_2,\gamma} \leq x)| \rightarrow 0. 
	\]
\end{theorem}

Theorems 1 and 2 show that we can use $\sum_{j=1}^{J_2} \lambda_{J_2,j}\chi_{1j}^2$ to approximate the asymptotic null distribution of $T_{n,J_2,\gamma}$ for both fixed and diverging $J_2$. The corresponding p-value can be calculated using the method of \citet{Davies1980}.

To show the asymptotic normality of $T_{n,J_2,\gamma}$ under the null, we need to impose the following assumption, which substitutes for specifying an explicit relationship between $J_2$ and $n$.

\begin{assumption}
	\label{asm5}
	$tr^2(R\Sigma_g)^2/tr(R\Sigma_g)^4 \rightarrow \infty$ and $tr(R\Sigma_g)^2 \rightarrow \infty$ as $n$ and $J_2 \rightarrow \infty$. 
\end{assumption}

\begin{theorem}
	\label{thm3}
	Under the null hypothesis $H_0$ in (\ref{h0}) and assume Assumptions 1--5 hold, 
	\[
	\sigma_{n,0}^{-1}\{ T_{n,J_2,\gamma} - tr(R \Sigma_s)\} \rightarrow N(0,1), 
	\]
	in distribution as $n$ and $J_2 \rightarrow \infty$.
\end{theorem}

We now study the interplay between the adverse effect of reduced sample size on power and the beneficial effect of variable selection afforded by sample splitting, under the local alternative $\mathscr{L}_{\beta}$,
$
\mathscr{L}_{\beta}=\left\{\Delta_{\beta}^TR\Sigma_gR\Delta_{\beta}=o\{n^{-1}tr(R\Sigma_g)^2\}\ \text{and} \ \{\mathcal{G}^{-1}(G_i^T\beta)\}^2 = O(1)\right\},
$
where $\Delta_{\beta} = E\{\mathcal{G}^{-1}(G_i^T\beta)G_i\}$.  

\begin{theorem}
	\label{thm4}
	Under the local alternative $\mathscr{L}_{\beta}$ and assume Assumptions 1--5 hold, 
	\[
	\sigma_{n,1}^{-1}\{T_{n,J_2,\gamma} - tr(R \Sigma_s)- \mu_{n,\beta}\} \rightarrow N(0,1),
	\] 
	in distribution as $n$ and $J_2 \rightarrow \infty$, where $\mu_{n,\beta} = tr(R\Xi_{\beta})+(n-1)\Delta_{\beta}^TR\Delta_{\beta}$, $\sigma_{n,1}^2 = \{2tr(R \Sigma_s + R\Xi_{\beta})^2\}\{1+o(1)\}$, and $\: \Xi_{\beta} = E\left[\{\mathcal{G}^{-1}(G_i^T\beta)\}^2G_iG_i^T\right]$. 
	
\end{theorem}

Theorem \ref{thm4} reveals that power of $T_{n,J_2,\gamma}$ under $\mathscr{L}_{\beta}$ is determined by 
$\text{SNR}_n(\beta) = \mu_{n,\beta}/\sigma_{n,1}$,
where $\text{SNR}_n(\beta)$ can be interpreted as signal-to-noise ratio following \citet{Guo2016}.
As detailed in the Supplementary Material, \\
$\mu_{n,\beta}= \sum_{j=1}^{J_2}r_jE\{\mathcal{G}^{-1}(G_i^T\beta)g_{ij}\}^2 + (n-1)\sum_{j=1}^{J_2}r_jE^2\{\mathcal{G}^{-1}(G_i^T\beta)g_{ij}\}$ and \\
$\sigma_{n,1}^2 = \{\sigma_{n,0}^2 + 2tr(R\Xi_{\beta})^2 +4tr(R\Sigma_s R\Xi_{\beta})\}\left\{1+o(1)\right\}$,  
where \\
$\sigma_{n,0}^2 = 2\sum_{j=1}^{J_2}\sum_{k=1}^{J_2} r_jr_k E^2(g_{ij}g_{ik}\epsilon_{i}^2)\{1+o(1)\}$, $tr(R\Xi_{\beta})^2=\sum_{j=1}^{J_2}\sum_{k=1}^{J_2} r_jr_kE^2\left[g_{ij}g_{ik}\{\mathcal{G}^{-1}(G_i^T\beta)\}^2\right]$, and 
$tr(R\Sigma R\Xi_{\beta}) =  \sum_{j=1}^{J_2}\sum_{k=1}^{J_2} r_jr_kE(g_{ij}g_{ik}\epsilon_{i}^2)E\left[g_{ij}g_{ik}\{\mathcal{G}^{-1}(G_i^T\beta)\}^2\right]$. 

Now define $T_{2n,J} = 2n\sum_{j=1}^JS_j^2$ as the test statistic calculated based on the {\it whole} sample of size $2n$ but {\it without} variable selection and  assuming $R=I$. In this case, $G_j \in \mathbb{R}^{ 2n \times 1}$ and $\beta \in \mathbb{R}^{ J \times 1}$ for calculating $S_j$. The signal-to-noise ratio corresponding to $T_{2n,J}$ is
$\text{SNR}_{2n}(\beta) = \mu_{2n,\beta}/\sigma_{2n,1}$. And \\
$\mu_{2n,\beta} = \sum_{j=1}^JE\{\mathcal{G}^{-1}(G_i^T\beta)g_{ij}\}^2 + (2n-1)\sum_{j=1}^JE^2\{\mathcal{G}^{-1}(G_i^T\beta)g_{ij}\}$, \\
$\sigma_{2n,1}^2 = \{\sigma_{2n,0}^2 + 2tr(\Xi_{\beta,J})^2 +4tr(\Sigma_{s,J} \Xi_{\beta,J})\}\left\{1+o(1)\right\}$,
where \\
$\sigma_{2n,0}^2 = 2\sum_{j=1}^J\sum_{k=1}^J E^2(g_{ij}g_{ik}\epsilon_{i}^2)\left\{1+o(1)\right\}$, $tr(\Xi_{\beta,J})^2=\sum_{j=1}^J\sum_{k=1}^J E^2\left[g_{ij}g_{ik}\{\mathcal{G}^{-1}(G_i^T\beta)\}^2\right]$, and $tr(\Sigma_{s,J} \Xi_{\beta,J}) =  \sum_{j=1}^J\sum_{k=1}^J E(g_{ij}g_{ik}\epsilon_{i}^2)E\left[g_{ij}g_{ik}\{\mathcal{G}^{-1}(G_i^T\beta)\}^2\right]$. 

To provide additional insights on power comparison, assume $pr(\mathcal{M} \supset \mathcal{M}^*) \rightarrow 1$ as $n \rightarrow \infty$; this assumption can be fulfilled by existing variable selection algorithms \citep{Fan2008,Li2012,Zhang2017}. Comparing $\mu_{n,\beta}$ with $\mu_{2n, \beta}$, it is not surprising that sample size reduction is the primary cause of power loss for a sample splitting-based method. However, the expressions for  $\sigma_{n,1}^2$ and $\sigma_{2n,1}^2 $ show that the first two terms are non-negative, and each term is a summation over $J_2$ and $J$ variants, respectively, for  $\sigma_{n,1}^2$ and $\sigma_{2n,1}^2 $. Because $J_2 \leq J$, noise-filtering in the training sample $D_{n,1}$ thus can reduce variance of the test statistic calculated in the testing sample $D_{n,2}$. 
Because SNR is ratio of $\mu$ over $\sigma$, $\text{SNR}_n(\beta)$ can be larger than $\text{SNR}_{2n}(\beta)$, and tests based on $T_{n,J_2,\gamma}$ can be more powerful than $T_{2n,J}$. 
The use of weights derived from $D_{n,1}$ can further compensate the efficient loss due to reduced sample size in $D_{n,2}$. Simulation studies in the next section show that, even if sure-screening fails in $D_{n,1}$, the sample splitting approach can have comparable power with the methods of \citet{Wu2019,Guo2016} applied to the full-sample without variable selection.

\section{Simulation studies}
\label{simulation}

\subsection{Simulation designs}
To evaluate the performance of $T_{dc}$ and compare it to tests proposed by \citet{Guo2016} and \citet{Wu2019}, we consider two simulation designs. Design one simulates $G$, while design two builds upon real genetic data from applications. Design one considers sample size of $2n=200$ or $1500$ and dimension $J \in \{10,50,200,400,1000,4000\}$. It generates $G$ based on a multivariate normal distribution with mean vector $0$ and (autoregressive) correlation matrix $\Sigma_g=\{\rho^{|i-j|}\}_{J \times J}$, where $\rho=0.2,0.5,0.8$, and $i$ and $j=1,\ldots,J$. For simulation design two, $G$ comes from two applications, where $2n=1409$ and $J=3754$ SNPs, and $2n=71$ and $J=4088$ gene-expression levels, respectively.  For a more streamlined presentation, we present simulation results of design two in Section \ref{application}, along with application results.

To implement $T_{dc}$, we let $r_j=\hat{\beta}_j^{\gamma-2}$ ($j=1, \ldots, J_2$) and $\Gamma = \{2,4,6,42\}$ to first obtain $T_c$ of (\ref{eq:Tc}). We then use $m=10$, $50$ or $100$ to derive the more stable $T_{dc}$ of (\ref{eq:Tdc}), and also to study the effect of $m$ on the performance of $T_{dc}$. For fair method comparison, we choose $\Gamma = \{2,4,6,42\}$ to be aligned with $\Gamma_{\text{aSPU}} = \{1,2,3,4,5,6, \infty\}$, studied and recommended by the authors of the aSPU test \citep{Wu2019}; $42$ in $\Gamma$ is to mimic $\infty$ in $\Gamma_{\text{aSPU}}$.  \citet{Wu2019} has also noted that $6$ ``often suffices and that the performance of the aSPU test is robust to such a choice'', which we observed for $T_{dc}$ in our studies (results not shown).  

For completeness, we also study the performance of the individual $T_1$ and $T_{\gamma}$'s ($\gamma=\{2,4,6,42\}$), but present the corresponding results in the Supplementary Material. The numbers of simulation replicates are $10^6$ for evaluating type I error control and 500 for power, and additional simulation design details are provided below when appropriate. 
\subsection{Type I error}
\label{type_I}

Methods applied to binary outcomes often have worse performance than normally distributed traits. Thus, we generate $Y$ based on a logistic regression with $\beta=0$, and without loss of generality, intercept equals to one and with no other covariates. For type I error evaluation, the variable selection procedure and the value of $J$ are not critical.
Thus, we choose $J_2=J$, regress simulated $Y$ on each of the $J_2$ simulated variants in $D_{n,1}$, and obtain the corresponding $\hat \beta_j$. We then perform the high dimensional polygenic association testing in $D_{n,2}$. 

Table \ref{table:1} shows the empirical test sizes of $T_{\gamma}$'s, $T_c$, and $T_{dc}$ for $2n=200$, $m=10,50,100$, $\rho=0.5$, and nominal $\alpha$ values of 0.05, 0.01, $10^{-3}$, and $10^{-4}$, and {Tables S1 and S2} in the Supplementary Material show results for $\rho=0.2$ and $0.8$, respectively; results for $2n=1500$ are more accurate thus not shown. Here, the distributions of $T_{\gamma}$'s are approximated by the weighted linear combination of independent $\chi_1^2$ distributions as specified in Theorem 2, and the distributions of $T_c$ and $T_{dc}$ by the standard Cauchy distribution. 

\begin{table}
	\centering
	\def~{\hphantom{0}}
	\caption{\label{table:1} Empirical test sizes for seven test statistics. Sample size $2n=200$, and autoregressive model $AR(1,\rho)$ with $\rho=0.5$ for correlation between the $J_2$ variants. One-time 50\%-50\% sample splitting for the first six methods, $m=10,50,100$ times sample splitting for the proposed $T_{dc}$.}\medskip 
	\resizebox{0.9\textwidth}{!}{
		\begin{tabular}{cccc cccc ccc}
			$J_2$ &$\alpha$ &$T_1$ & $T_2$ & $T_4$ & $T_6$ & $T_{42}$  & $T_c$& $T_{dc,10}$ &$T_{dc,50}$ & $T_{dc,100}$\\\\ 
			
			10& 5\% & 4.9975 & 4.7951 & 4.8761 & 4.9097 & 4.9130 & 5.1774&5.2651&5.2281 &4.8201\\   
			& 1\% & 0.9640 & 0.8771 & 0.8927 & 0.8999 &0.9250 &0.9510 & 0.9614&1.0518 &1.0082\\
			& 0.1\%& 0.0873 & 0.0768 & 0.0695 &0.0700 & 0.0746& 0.0789&0.0951&0.0854 &0.0730\\  
			& 0.01\% &0.0078 & 0.0063 & 0.0052 & 0.0051 & 0.0046 &0.0088&0.0081&0.0092 &0.0096\\\\ 
			
			50& 5\%    & 4.9503 & 4.7245 & 4.7715 &4.8447 & 4.9615 & 5.1745 &5.3395&5.3038 &5.2115\\   
			& 1\%    & 0.9701 & 0.8750 &0.8912  &0.9110 & 0.9339 & 0.9449&0.9360&0.9280 &0.9033\\
			& 0.1\%  & 0.0961 & 0.0772 &0.0769   &0.0789 & 0.0767 & 0.0825&0.0698&0.0604 &0.0598\\  
			& 0.01\% & 0.0087 & 0.0064 &0.0059  &0.0049 & 0.0056 & 0.0062&0.0041&0.0067&0.0110\\\\ 
			
			200& 5\% & 4.9820 & 4.6781 & 4.7274 & 4.8011 & 4.9341 & 5.1217 & 5.3721&5.4933&5.5147\\   
			& 1\% & 0.9921 & 0.8743 & 0.8951 & 0.8961 & 0.9172 & 0.9517&0.9408 &0.9416&0.9374\\   
			& 0.1\%& 0.0960 & 0.0814 & 0.0780 &0.0775 & 0.0775 & 0.0815& 0.0705&0.0656&0.0679\\  
			& 0.01\% & 0.0116 & 0.0070 & 0.0067 &0.0064 & 0.0060 & 0.0069&0.0042&0.0156&0.0263 \\\\ 
			
			400& 5\% & 4.9763 & 4.6932 & 4.7571 & 4.8307 & 4.9550 & 5.1717 &5.3691&5.5539&5.6548\\   
			& 1\% & 0.9848 & 0.8886& 0.8991 & 0.9161 & 0.9230 & 0.9518&0.9375&0.9527&0.9679\\   
			& 0.1\%& 0.0972 & 0.0786& 0.0830& 0.0790 & 0.0767 & 0.0824 &0.0724&0.0624&0.0740\\  
			& 0.01\% & 0.0102 &0.0061& 0.0070&0.0053 &0.0049 &0.0067&0.0053&0.0150&0.0245\\\\ 
			
			1000& 5\% &4.9887 & 4.6689 & 4.7323 & 4.7755 & 4.9269 & 5.1299 & 5.4092&5.7421& 5.1065\\
			& 1\% &1.0117 &0.8923 & 0.8829 & 0.08865 & 0.9227 & 0.9312 & 0.9405&0.9734&0.8623\\
			& 0.1\%&0.0967& 0.0754&0.0808 & 0.0755 &0.0798 &0.0821 &0.0763&0.0604&0.0564\\  
			& 0.01\% &0.0100&0.0073&0.0082 & 0.0060 &0.0059 & 0.0074 & 0.0047&0.0087&0.0173\\\\ 
		\end{tabular}
	}
	\label{tablelabel}
\end{table}

Table \ref{table:1} shows that the empirical type I error rate of $T_{dc}$ is controlled at or below the nominal $\alpha$ level when $m=10$ and $50$, considering Monte Carlo error. However, the empirical type I error rate is slightly inflated for larger $J_2$ and stringent $\alpha$ level ($\alpha=10^{-4}$) when $m=100$. To better understand this inflation problem, we provide the summary statistics of the empirical $\alpha$ from the $m=100$ sample splits in Table S3 when $J_2=400$ and  $1000$. Results show that the test size of $T_{c}$ is accurate and stable across the 100 sample splits. Thus, the inflation stems from the double Cauchy combination step. The accuracy of the Cauchy approximation for large $m$ has been studied in Theorem 2 by \citet{Liu2019}. \citet{Liu2019} showed that, to obtain accurate p-value approximation, $m$ should be bounded by $(t_{\alpha})^{c_0}$, where $t_{\alpha}$ is the upper $\alpha$-quantile of the standard Cauchy distribution and $0<c_0<1/2$. When $\alpha = 10^{-4}$, $t_{\alpha} = 3183$, thus $t_{\alpha}^{1/2} = 56.4$ provides an upper bound of $m$. Although this theoretical result is under certain conditions on the correlation matrix of the p-values to be combined and not accurate across all scenarios \citep{Liu2019}, it helps the understanding of the approximation error when $m=100$. 

In the above simulation studies and later in applications, the p-value approximation for the individual $T_{\gamma}$ is based on $\sum_{j=1}^{J_2} \lambda_{J_2,j}\chi_{1j}^2$ as shown in Theorem 2. Table S4 shows that the normal approximation given in Theorem 3, however, is not adequate for stringent $\alpha$ levels when $\gamma>2$ and using the simulation parameter values considered here. Thus, we recommend the use of the $\sum_{j=1}^{J_2} \lambda_{J_2,j}\chi_{1j}^2$ approximation in practice. Consistent with previous reports, tests of \citet{Wu2019} and \citet{Guo2016} based on asymptotic approximations are not accurate (Table S4). Thus, we use parametric bootstrap, as recommended by the authors, with $10^3$ replicates to evaluate power of these methods for fair comparison.

\subsection{Power}\label{power}
Similar to the type I error evaluation above, here we also focus on the more difficult case of analyzing binary outcomes than normally distributed traits. We generate $Y$ based on logistic models with different proportions of nonzero regression coefficients, varying from $0.1\%$, $1\%$, $5\%$, to $10\%$ for the $J$ variants. We assume the indices of the nonzero $\beta_j$'s to be uniformly distributed in $\{1,\ldots,J\}$. We consider three different scenarios of the sign of nonzero $\beta_j$'s: randomly specify the nonzero $\beta_j$'s to be half positive and half negative, all the nonzero $\beta_j$'s are positive, and all are negative. Results below focus on power comparison between the proposed $T_{dc}$ test and the methods of \citet{Guo2016} and \citet{Wu2019}, which are applied to the whole sample and without variable selection. Results of the original polygenic risk score test, $T_1$, are shown in Figure S1.

To better delineate the factors influencing power, we consider three study scenarios. In all three scenarios, the weights inferred from the training sample $D_{n,1}$ are leveraged to construct the $T_{dc}$ test statistic using the testing sample $D_{n,2}$. 

(I), Oracle: $\mathcal{M} = \mathcal{M}^*$. This is the `best' case scenario for $T_{dc}$, where the selection step applied to $D_{n,1}$ identifies all and only truly associated variants; the estimated weights however may not be optimal. This study is to show power gain of $T_{dc}$, despite the reduction of sample size, as compared to the methods without variable selection. 

(II), $J_2=J$: $\mathcal{M} = \{G_1,\ldots, G_J\}$. This is the `worst' case scenario for $T_{dc}$, where the selection step fails completely at filtering out non-signals; the estimated weights however may be informative. This scenario is tailored for studying power loss of $T_{dc}$ due to sample size reduction as compared with methods without sample splitting, while also demonstrating the benefits of leveraging the weights inferred from $D_{n,1}$ for associate testing using $D_{n,2}$.

(III), Variable Selection: $\mathcal{M}$ is estimated after variable screening. This study investigates the impact of accuracy of variable selection on power of $T_{dc}$ as compared to the methods without variable selection.

For variable selection, we considered the DCSIS method of \citet{Li2012} (DCSIS) and SIS of \citet{Fan2008}, because these methods require less assumptions than e.g. {ElasticNet \citep{Zou2005}} for the property of sure screening to hold \citep{buhlmann2014high}. The implementation of DCSIS and SIS also require the specification of $J_2$. 
In our power simulation study, $2n=1500$ and $J=4000$, and we choose $J_2 = 2n$, which is more conservative than $2n/log(2n)$ recommended by the authors. 

\begin{figure}[!htp]
	\centering
	\includegraphics[width=.8\textwidth]{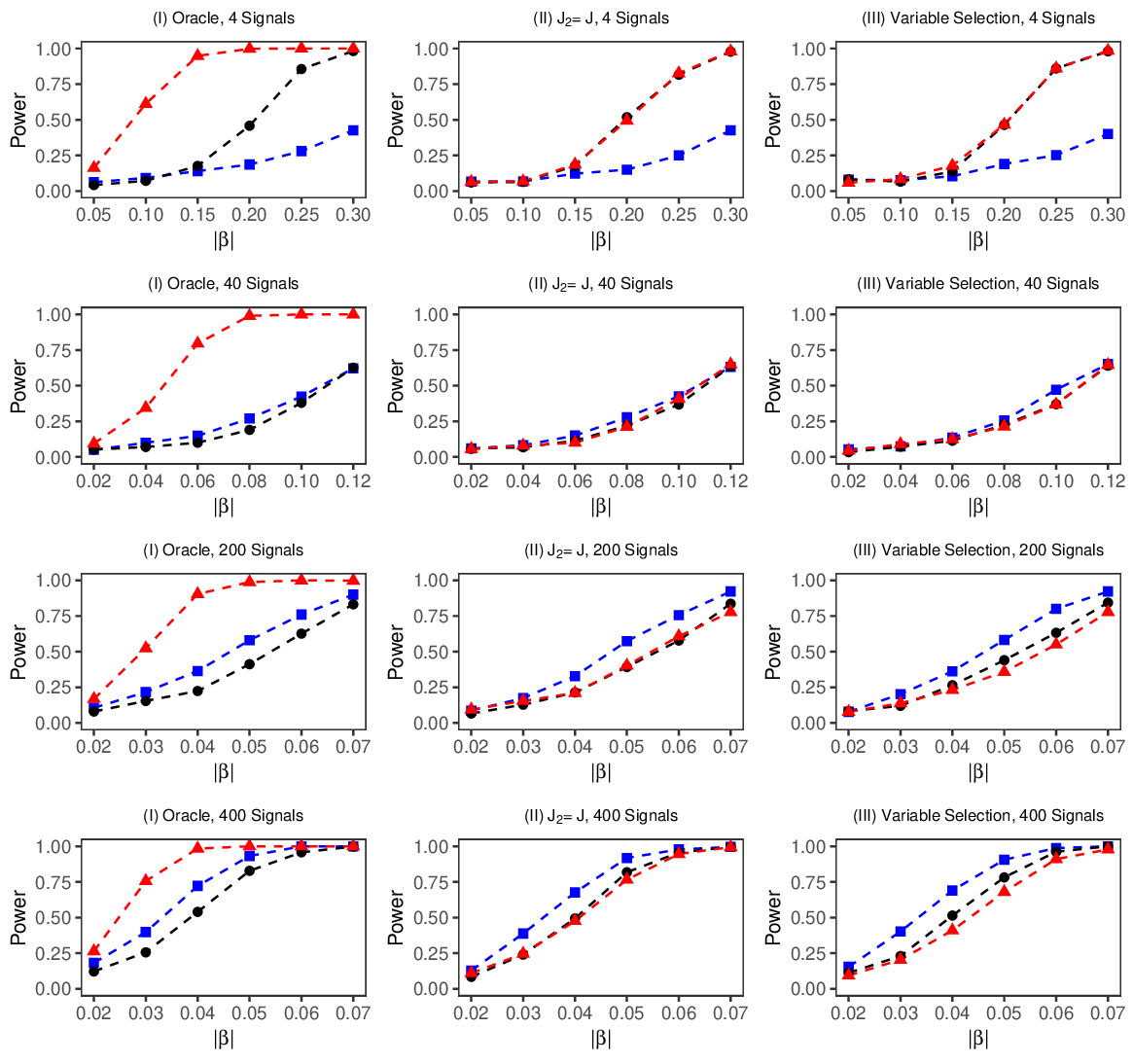}
	\caption{\label{fig:power} Power comparison of the proposed test $T_{dc}$ (red triangle), the method of \citet{Guo2016} (blue square), and the method of \citet{Wu2019} (black circle), for the three study scenarios (I), (II) and (III). Sample size $2n=1500$, and the total variant $J=4000$ among which 0.1\% (row 1), 1\% (row 2), 5\% (row 3), and 10\% (last row) are truly associated.}
\end{figure}

Because all three methods, the proposed $T_{dc}$ test and the methods of \citet{Guo2016} and \citet{Wu2019}, incorporate $S^2_j$'s across variants, we expect them to be robust to the direction/sign of $\beta_j$'s. This is confirmed by {Figure S2}, which also shows that results are qualitatively similar between the two variable selection methods, DCSIS of \citet{Li2012} and SIS of \citet{Fan2008}. Thus, below we only present the simulation results when the nonzero $\beta_j$'s are half positive and half negative, and DCSIS is the variable selection method.  Figure \ref{fig:power} shows the results for $2n=1500$ and $J=4000$, reflecting the values observed in the real dataset studied in Section \ref{application}; $\rho=0.5$ for correlation between the $J$ variants, $J_2=1500$ for variable selection, and $m=10$ for repeated 50\%-50\% sample splitting to construct $T_{dc}$. 

For scenario (I), the first column of Figure \ref{fig:power} shows that the proposed $T_{dc}$ test has substantial power gain, attributed to noise filtering despite of the reduction in sample size for associate testing (and using the estimated weights), as compared to the methods of \citet{Guo2016} and \citet{Wu2019}. 

For scenario (II), the second column of Figure \ref{fig:power}  shows that the anticipated power loss of $T_{dc}$ due to sample splitting can be compensated by leveraging the weights inferred from $D_{n,1}$, as compared with the methods of \citet{Guo2016} and \citet{Wu2019}, which use the full sample; recall that $J_2=J$ for $T_{dc}$, meaning the variable selection step completely failed at selecting relevant variants. For the sparse alternative case, scenario (II) 4 signals, $T_{dc}$ displays comparable power with the method of \citet{Wu2019}, while both are substantially more powerful than the method of \citet{Guo2016}. For the other alternatives considered in this scenario, all three methods have comparable power with the method of \citet{Guo2016} having slightly higher power. Overall, the proposed $T_{dc}$ test is most robust to the different alternatives considered here, and it is also computationally efficient which we discuss in Section \ref{discussion}. 

For the more realistic scenario (III), interestingly the results are similar to those of scenario (II). This suggests that while the variable selection step filters out noise, it also filters out some (weak) signals; the sure screening property requires that the nonzero regression coefficients must be sufficiently large \citep{buhlmann2014high, Fan2008}. In addition to DCSIS and SIS, we also evaluated other selection methods such as ElasticNet, but results are similar especially for weak signals. 

We emphasize that $J_2=2n=1500$ is fixed across all the alternatives considered  here, and power of $T_{dc}$ shown in Figure \ref{fig:power} can be improved by considering a smaller $J_2$ for sparse but relatively strong signals (e.g.\ 4 or 40 signals out $J=4000$ variants). Indeed, applications and additional simulation studies in Section \ref{application} demonstrate the advantages of the proposed method for certain alternatives. Power can be further improved by using $m=50$ instead of $10$ (Figure S3). In general, a larger $m$ leads to a heavier penalty paid for multiple hypothesis testing; the effective number of tests, however, does not go up linearly with respect to $m$ because of the inherent correlation between the different sample splits. On the other hand, a larger $m$ leads to better variable selection in stage 1, particularly when the signals are weak. The resulting improved efficiency thus compensates the power loss due to multiple hypothesis testing. Overall, performance of $T_{dc}$ is robust to the choice of $m$ (Figure S3).

Simulation results so far have focused on the 50\%-50\% sample splitting proportion. We have also investigated 33\%-67\% (Figure S4) and  67\%-33\% (Figure S5) sample splitting. Results in Figures S4 and S5 show that the overall power of $T_{dc}$ is not very sensitive to the proportion.  However, the 33\%-67\% sample splitting has slightly increased power for the scenarios considered here. This is consistent with the literature \citep{Barber2019}, where it has been noted that uneven sample splitting, with more subjects assigned to the testing sample, can increase power as compared to even sample splitting.

\section{Application and additional simulation studies}\label{application}
\subsection{Cystic fibrosis data}\label{cfd}
We apply the proposed $T_{dc}$ test and the methods of \citet{Guo2016} and \citet{Wu2019}, as well as
$T_1$, the original polygenic risk score test, to the cystic fibrosis data introduced in \citet{David2015}. This dataset consists of $2n=1409$ independent individuals from Canada with cystic fibrosis on whom lung functions have been measured. Of interest is the association between lung function and a set of $J=3754$ genetic variants, which are the constituents of the apical plasma membrane. These are candidates for association with cystic fibrosis but selected unsupervised based on biological hypothesis alone \citep{sun2012}.

To implement the proposed $T_{dc}$ test, we first randomly divide the $1409$ individuals into two subsets, $D_{n_1}$ and $D_{n_2}$, where $n_1:n_2=409:1000$, $n_1:n_2=705:704$, and $n_1:n_2=1000:409$. As in the simulation studies, we define $r_j=\hat{\beta}_j^{\gamma-2}, j=1, \ldots, J_2$, and $\Gamma = \{2,4,6,42\}$, and we apply the variable selection method DCSIS \citep{Li2012} and let $J_2=n_2$, the sample size of the testing sample. Because the approximation of the asymptotic distribution of $T_{dc}$ requires a positive definite matrix estimate of  $\Sigma_g$ in $D_{n_2}$, we use the algorithm proposed by \citet{Rothman2012}, with the tuning parameter selected by 5-fold cross-validations.  

Using this application dataset, we first re-evaluate the accuracy of $T_{dc}$ by simulating $Y$ independently of the observed $G$ for the $n_2$ individuals in  $D_{n_2}$, where $y_i=1+\varepsilon_i$, $i=1,\ldots,n_2$, and $\varepsilon_i$ follows the standard normal distribution. For variable selection and weight estimation, simulating $Y$ for the $n_1$ individuals in $D_{n_1}$ is an obvious approach. However, to see how potentially non-random variable selection and weight estimation adversely affect the type I error control of the proposed method, we used the real data, both $Y$ and $G$, of the $D_{n_1}$ {\it training} sample; note that $Y$ in the $D_{n_2}$ {\it testing} sample is simulated. 

Table \ref{table:3} shows that the empirical $\alpha$ level of $T_{dc}$ remains well controlled when $m=10$ or 50, but is slightly inflated when $m=100$. This result is similar to that based on simulated multivariate normal predictors in Section \ref{type_I}, where we showed that the upper bound for $m$ is 56. In general, smaller $m$ leads better type I error control, but larger $m$ provides more inference stability. Thus, we will use $m=50$ in our real data analyses.  

\begin{table}
	\centering
	\caption{\label{table:3} Empirical test sizes for $T_{dc}$ based on simulated outcome values but real genetic data from the two application datasets. For the cystic fibrosis application data, $2n = 1409$ and $J = 3754$, and for the Riboflavin application data, $2n = 71$ and $J = 4088$. $n_1$ is the sample size for the training sample and $J_2$ is the number of selected variants for association testing in the testing sample.}\medskip 
	\resizebox{0.4\textwidth}{!}{
		\begin{tabular}{cccc ccc }  
			Cystic Fibrosis &$J_2$ & $n_1$ &\diagbox[width=2.5em]{$\alpha$}{$m$} & 10 & 50 & 100  \\\\
			&409 & 1000  &5\%  & 5.0021 & 4.8969 & 4.8873  \\   
			&&&1\%  & 0.9981 & 1.1371 & 1.1389  \\
			&&&0.1\% & 0.1107 & 0.1067 & 0.0869 \\ 
			&&&0.01\% & 0.0084 & 0.0137 & 0.0207\\\\
			
			&704  & 705 &5\%  & 4.9311  & 4.6359  & 4.7072 \\   
			&&&1\%  & 0.9847  & 0.9619  & 1.1109  \\
			&&&0.1\% & 0.1034  & 0.0843  & 0.0897  \\ 
			&&&0.01\% & 0.0077  & 0.0149  & 0.0274 \\\\
			
			&1000  & 409 &5\%  & 4.8180  & 4.7277  & 5.8078 \\   
			&&&1\%  & 0.9572  & 1.0699  & 2.3768 \\
			&&&0.1\% &0.1001  & 0.1017  & 0.1891 \\ 
			&&&0.01\%  &0.0061& 0.0170  & 0.0354 \\\\        
			
			
			Riboflavin &36 & 35  &5\%  & 4.6538  &  3.5662  & 3.3164\\   
			&&&1\%  &  0.7951  &  0.8661  &1.1555\\
			&&&0.1\% & 0.1243 &   0.1120 &0.1117\\ 
			&&&0.01\% &  0.0093  & 0.0096 &0.0116\\ 
		\end{tabular}
	}
	\label{tablelabel}
\end{table}

In the absence of oracle knowledge of true association, application results focus on discussing the range of p-values of all methods. The empirical p-values are 0.0985 and 0.0727, respectively, for the methods of \citet{Guo2016} and \citet{Wu2019}, based on $10^4$ bootstrap samples applied to the whole sample. For $T_1$, we randomly split the whole sample to the $D_{n_1}$ and $D_{n_2}$ subsets, independently 100 times, to obtain 100 different $p_{1}$'s, the $T_1$-based p-values. The histogram of $p_{1}$'s for $n_1:n_2=409:1000$ is shown in Figure \ \ref{fig:cf}. 

For $T_{dc}$, we also randomly split the whole sample to two subsets, but independently $100\times 50$ times, and use a sequence of $m=50$ repeated sample splits to obtain 100 $p_{dc}$'s, the $T_{dc}$-based p-values. The histogram of $p_{dc}$'s for $n_1:n_2=409:1000$ is shown in Figure \ref{fig:cf} in blue. Results clearly show that the proposed repeated sample splitting strategy leads to a much more stable inference than the one-time-only sample splitting approach: $p_1$ ranges from 0.0019 to 0.9446, while $p_{dc}$ ranges from 0.003 to 0.101 with a mode of around 0.05. For completeness, we also provide the summary statistics of the 100 $p_{dc}$'s in Table \ref{table:2}. 

In an effort to study the behaviour of $T_{dc}$ in depth, we considered three different sample splitting proportions, 33\%-67\%, 50\%-50\% and 67\%-33\%, and obtained 100 p-values for each proportion to demonstrate the improved inference stability of $T_{dc}$ as compared with $T_1$. However, a practical question arises: What would be the reported p-value for this application? This could be randomly drawn from the set of p-values, but we recommend the medium of the 100 p-values from the 50\%-50\% sample split for a conservative estimate; 33\%-67\% can increase power as compared to 50\%-50\%, assuming sufficient total sample size \citep{Barber2019}. Overall, considerable variation remains in this application, suggesting that the signals are too weak or sample size is not sufficient.

\begin{table}
	\centering
	\caption{\label{table:2} Summary of p-values of the proposed $T_{dc}$ applied to two real application datasets based on different $n_1$-$n_2$ sample splits, with $m=50$ for constructing $T_{dc}$ and repeatedly 100 times. The total sample size $2n=1409$ for cystic fibrosis data and $2n=71$ for riboflavin data, and the total genetic variants $J=3754,\ 4088$, respectively. We choose $J_2=n_2$, the sample of size of the testing sample for all scenarios}\medskip 
	\resizebox{0.9\textwidth}{!}{
		\begin{tabular}{ccc cccccc }
			&	$n_1$&$n_2$ & Minimum & 1st Quantile & Median & Mean & 3rd Quantile & Maximum  \\\\
			Cystic Fibrosis	&409 & 1000&  0.003 & 0.030 & 0.045 & 0.046 & 0.061 & 0.101    \\ 			
			&705 &704   & 0.020 & 0.067  & 0.079    &0.086   &0.105   &0.195 \\
			&1000&409  & 0.003  & 0.045  & 0.059  & 0.057 & 0.067 & 0.104  \\\\
			
			Riboflavin &  35 & 36 &  $5.551 \times 10^{-17}$  &  $1.665 \times 10^{-16}$  & $1.665 \times 10^{-16}$  & $2.520 \times 10^{-16}$ & $2.359 \times 10^{-16}$ & $4.496 \times 10^{-15}$  \\			
			
		\end{tabular}
	}
	\label{tablelabel}
\end{table} 

\subsection{Riboflavin data}
In this application, the outcome of interest ($Y$) here is the standardized riboflavin (B2) production rate, measured on $2n=71$ independent individuals, and the predictors ($G$) are standardized gene expression levels of $J=4088$ genes. {The dataset is freely available in the R package hdi, and has been used for studying variable selection \citep{buhlmann2014high} and constructing valid confidence interval after model selection \citep{shi2020statistical}.} 

Similarly to the application above in Section \ref{cfd}, we first use the real $G$ but simulated $Y$ to re-evaluate the type I error control of $T_{dc}$.  Results in Table \ref{table:3} shows that $T_{dc}$ remains accurate in this setting. Considering the total sample size of $n=71$, the implementation of $T_{dc}$ in this study only used the 50\%-50\% sample splitting proportion, where $n_1:n_2=35:36$ and $J_2=36$ for variables selected using DCSIS \citep{Li2012}. 

We then apply the three methods to the real data for method comparison. The empirical p-values are $0.028 $ and $5.0 \times 10^{-5}$, respectively, for the methods of \citet{Guo2016} and \citet{Wu2019} based on $10^5$ bootstrap samples. In contrast, the p-value of $T_{dc}$ is in the range of $10^{-16}$ based on $m=50$. 

To show the stability of our inference, similar to Section \ref{cfd}, we also perform sample splitting independently $100 \times 50$ times to obtain 100 $p_{dc}$'s. The summary statistics of the 100 $p_{dc}$'s are shown in Table \ref{table:2}. The maximum $p_{dc}$ is $4.496 \times 10^{-15}$, suggesting better performance of $T_{dc}$ as compared to the other two methods. For completeness, we also compare the performance between $T_{dc}$ and $T_1$. Results in Figure S6 show that the proposed repeated sample splitting not only provides robustness but also improves power; the range of $p_1$ is $[7.8 \times 10^{-8}, 0.027]$ as compared with $[5.551 \times 10^{-17}, 4.496 \times 10^{-15}]$ of $p_{dc}$.

\begin{figure}[!htp]
	\centering
	\includegraphics[width=.9\textwidth]{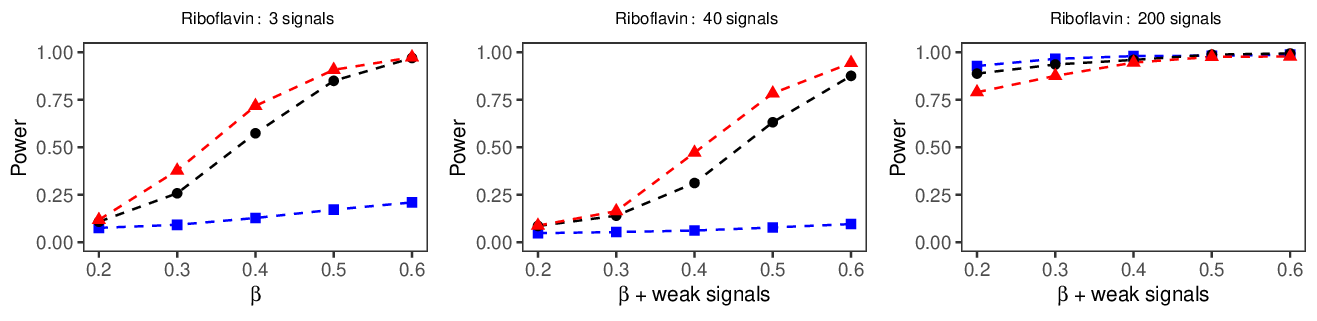}
	\caption{\label{fig:riboflavin} Power comparison of the proposed test $T_{dc}$ (red triangle), the method of \citet{Guo2016} (blue square), and the method of \citet{Wu2019} (black circle) based on the real gene expression data of the Riboflavin dataset. In total, $2n=71$ and $J=4088$; $J_2=n_2=36$ for variables selected using DCSIS.  Among the 4088 variants, 3 (left), 40 (middle), 200 (right) are truly associated. In each case, 3 signals are relatively strong with their $\beta_j$'s shown in the X-axis, and the remaining weak signals with their $\beta_j$'s randomly drawn from $N(0, 0.05)$.}
\end{figure}

To further demonstrate the reliability of the application result, we conduct additional power simulation study based on the real gene expression levels ($G$) of the 4,088 genes. To simulate $Y$, we consider the number of nonzero regression coefficients to be 3, 40 or 200. First, to best mimic the presumed underlying signal structure \citep{shi2020statistical}, we assume $\beta_j,\ j=1588,\ 3154$ and $4004$ to be nonzero and all the $\beta_j$s are positive with values ranging from 0.2 to 0.6 (Figure \ref{fig:riboflavin}). Next, for the scenarios where there are 40 and 200 signals in total, we further assume the first 37 and 197 $\beta_j$'s are nonzero, respectively; these $\beta_j$'s are generated from a normal distribution with mean 0 and variance 0.05, corresponding to weak signals. To implement $T_{dc}$, we use $n_1:n_2=35:36$ and $J_2=36$ for variables selected using DCSIS, the same set-up as that for type I error evaluation above. 

This simulation falls under scenario (III) considered in Section \ref{power}, where power study was based on both simulated $G$ and $Y$, and the corresponding power is shown in the right column of Figure \ref {fig:power}. Here, results in Figure \ref{fig:riboflavin} show that the method of \citet{Guo2016} (blue square) performs poorly when there are sparse strong signals (left plot) or sparse strong signals combined with some weak signals (middle plot). In either case, power of the proposed $T_{dc}$ test is appreciably bigger than that of the method of \citet{Wu2019} (black circle). To be consistent with the power study in Section \ref{power}, we note that power in Figure \ref{fig:riboflavin} is for $m=10$. Figure S7 shows that power of $T_{dc}$ can be slightly improved by using $m=50$; this is consistent with the results in Figure S3. 

\section{Discussion}
\label{discussion}

In the theoretical study, we did not consider the impact of estimating nuisance parameters $\beta_x$ and $\phi$, as we expect that the results would be similar when we impose stringent conditions on the design matrix $X$ and the relationship between $n$ and $q$ to ensure estimation accuracy of the nuisance parameters \citep{Guo2016}. In practice, we can estimate the nuisance parameters in the training sample and treat the estimates as known quantities to construct $T_{dc}$ in the testing sample. This approach has been recommended by \citet{Cher2018} for another study setting where the sample splitting strategy is used. 

The proposed $T_{dc}$ is computationally efficient, because a) p-value of each $T_{\gamma}$ is estimated by using the null distribution derived in Theorem 2, $\sum_{j=1}^{J_2} \lambda_{J_2,j}\chi_{1j}^2$, and b) p-value of $T_{c}$ across different $\gamma$ values is calculated using equation (\ref{eq:Tc}), and c) the final p-value of $T_{dc}$ across $m$ sample splits is obtained using equation (\ref{eq:Tdc}). For example, using a laptop with Apple M1 Chip with 8 GB unified memory, the computation time for the riboflavin data application took 6.3 seconds with $m=10$ and 31.7 seconds with $m=50$; the computation time scales linearly with respective to $m$, as expected. The computation cost mostly comes from variable selection, as the computation time is reduced to 2 and 10 seconds, respectively, for $m=10$ and 50 when using SIS instead of DCSIS for variable selection. 

For easy of implementation, we have developed a R package, DoubleCauchy, and released it at https://github.com/yanyan-zhao/DoubleCauchy. The package contains three main functions: i) the DoubleCauchy function to conduct the proposed $T_{dc}$ test, where the variable selection step can be based on DCSIS, SIS or ElasticNet, ii) the DoubleCauchyParallel function to further reduce the computation time of the DoubleCauchy function, if parallel computing resource is available, and iii) the AdapSide function to leverage additional information available, such as the functional importance measure of each of the variants analyzed. 

\section*{Acknowledgements}
We thank Dr.\ Lisa J.\ Strug and her lab for providing the cystic fibrosis data for application and helpful discussion. YZ is a trainee of the CANSSO-Ontario STAGE training program at the University of Toronto. This research is funded by the Natural Sciences and Engineering Research Council of Canada, the Canadian Institutes of Health Research, and the University of Toronto McLaughlin Centre.

\newpage
\bibliographystyle{ametsoc2014}\bibliography{paper-ref}
 
\end{document}